\documentclass[aps,pra,reprint,superscriptaddress,amsmath,amssymb,floatfix]{revtex4-2}

\usepackage{graphicx}  % Include figure files
\usepackage{bm}        % Bold math
\usepackage{tikz}      % Keep since you used it for diagrams

\usepackage[colorlinks=true,citecolor=blue,linkcolor=blue,urlcolor=blue]{hyperref}

\begin{document}

% 4. Title & Author Block
\title{Classical timing noise in gravity-mediated entanglement tests:\\
LOCC structure, metrological bounds, and visibility thresholds}

\author{Yaghmorassene Hebib}
% Optional: \email{hebibya@butte.edu} 
\affiliation{Butte College, Department of Physical Science, Oroville, California, USA}

\date{\today} % PRA usually prints the date

% 5. Abstract MUST go before \maketitle in REVTeX
\begin{abstract}
Table-top proposals to test gravity-mediated entanglement aim to distinguish coherent gravitational interactions from classical dephasing processes that generate identical phases on both interferometers. A particularly important contribution is platform-invariant timing noise, which can be accessed through optical-clock cross-spectra and frequency-transfer links. In this work we (i) derive the mapping from clock cross-spectra to an effective common-mode dephasing rate, (ii) show that the corresponding dynamical channel is LOCC and therefore unable to generate entanglement, and (iii) combine published clock and interferometer noise floors to obtain quantitative bounds on the ratio between common-mode and local dephasing rates in representative gravity-entanglement proposals. Across QGEM-, MAQRO-, and levitated-nanoparticle regimes we find a robust hierarchy $\gamma_{\rm com}/\gamma_{\rm loc}\sim10^{-9}$--$10^{-11}$, identifying platform-invariant timing noise as a negligible but calibratable background. We further derive visibility and Bell--CHSH thresholds in the presence of both common-mode and local dephasing, and illustrate the full calibration workflow with a simple synthetic example anchored to state-of-the-art clock metrology. Extended derivations and statistical tools are provided in the Supplemental Material.
\end{abstract}

\maketitle

%%%%%%%%%%%%%%%%%%%%%%%%%%%%%%%%%%%%%%%%%%%%%%%%%%%%%%%%%%%%%%%%%%%%%%%%%%%%%%%%%%%%%%%%%%%%%%%%%

\section{Introduction}

The prospect of testing whether gravity can mediate quantum entanglement has motivated a new class of tabletop proposals that place two mesoscopic masses into spatial superposition and monitor their mutual interaction through interference visibility or entanglement witnesses. Seminal analyses by Bose \textit{et al.}~\cite{Bose2017} and Marletto and Vedral~\cite{Marletto2017} established that, if entanglement is generated between two otherwise isolated systems, the mediating field cannot be described by a purely classical channel. This criterion underlies current experimental designs ranging from nanosphere interferometry (QGEM) to cryogenic space-based architectures (MAQRO) and levitated optomechanical platforms, which aim to probe the regime where gravitationally induced phase correlations become comparable to environmental decoherence.

A central challenge in such experiments is that classical dephasing processes can mimic or obscure the phases that would otherwise certify quantum-mediated interactions. Several theoretical frameworks describe these classical contributions. The classical-channel model of Kafri, Taylor and Milburn~\cite{Kafri2014} shows that any classical (commuting) gravitational interaction can be represented as a correlated dephasing channel incapable of generating entanglement. Stochastic-gravity approaches, notably those of Anastopoulos and Hu~\cite{Anastopoulos2015}, treat the metric as a fluctuating classical field and predict colored noise kernels that imprint correlated phases on both masses. Collapse models in the Di\'osi--Penrose tradition~\cite{Diosi2011,Penrose1996} assign an intrinsic decoherence rate proportional to the mass--density self-energy, again producing dephasing that does not generate coherent correlations. These theories differ in physical interpretation, but they share a common structure: they produce collective phase fluctuations that remain in the class of separability-preserving channels. For experimental efforts, the crucial question is therefore not the existence of such classical sectors---these are well understood---but their quantitative strength and how to calibrate them against the coherent gravitational interaction.

Despite extensive theoretical work, there is no operational framework for translating measurable timing fluctuations into quantitative bounds on the classical common-mode dephasing present in gravity-entanglement experiments. Existing models provide important conceptual insights but do not specify how to incorporate real metrological data, such as the cross-spectra measured between state-of-the-art optical clocks or frequency-transfer links. Conversely, the experimental literature on optical clocks and time-transfer networks has achieved record low frequency-correlation floors~\cite{Oelker2019,Bothwell2022,Takamoto2020}, but these results have not been connected to the specific geometries and timescales of gravity-mediated entanglement proposals. As a result, the community currently lacks a unified method to determine how much of the observed visibility loss in a two-mass interferometer is attributable to platform-invariant timing noise, and how much reflects local technical decoherence or coherent gravitational coupling.

The present work fills this gap by establishing a quantitative and fully operational connection between optical-clock metrology and gravity-mediated entanglement tests. First, we derive a direct mapping from clock cross-spectra to an effective common-mode dephasing rate that applies to any two-mass geometry relevant to existing proposals. Using the published low-frequency floors of leading optical-clock networks, this mapping provides numerical bounds on the strength of platform-invariant proper-time fluctuations. Second, we identify the dynamical channel generated by these fluctuations as a convex mixture of correlated local phase rotations, placing it squarely in the class of LOCC (local operations and classical communication) channels. This connection is conceptually known in abstract quantum-information settings, but here it is given concrete physical meaning: the platform-invariant proper-time sector is not only classical but also directly measurable and can be incorporated as a calibration contribution in visibility analyses. Third, by combining this calibration bound with representative decoherence budgets from QGEM-, MAQRO-, and levitated-nanoparticle platforms~\cite{RomeroIsart2011,Millen2020}, we show that the resulting common-mode dephasing rate is nine to eleven orders of magnitude smaller than typical local decoherence rates. This hierarchy identifies platform-invariant timing fluctuations not as an experimental limitation but as a negligible yet quantifiable background that should accompany all analyses of gravitational entanglement visibility.

Building on these results, we derive visibility and Bell--CHSH thresholds in the presence of both local and common-mode dephasing. The resulting conditions give practical entanglement-feasibility criteria for current and near-term platforms and illustrate how the classical timing floor enters entanglement-witness analyses. To demonstrate usage rather than validation, we also include a simple synthetic end-to-end example that illustrates how clock-based bounds on the common-mode rate can be combined with interferometric visibility fits to isolate local technical decoherence. All extended derivations, scaling relations, and statistical tools are provided in the Supplemental Material.

Taken together, these results provide a unified calibration framework for forthcoming gravity-entanglement experiments. They anchor the classical sector to measurable clock correlations, guarantee that this sector is separability-preserving, and quantify its magnitude relative to local decoherence. As increasingly precise interferometers and clock networks become available, deviations from the established hierarchy or residual dephasing beyond clock-predicted bounds would offer a clear diagnostic of new physical effects or uncharacterized environmental channels. The framework therefore supplies both a quantitative calibration baseline and a pathway toward the robust interpretation of future gravitational entanglement data.

\section{Proper-time fluctuations and common-mode dephasing}
\label{sec:proper_time}

Gravity-mediated entanglement proposals rely on the relative phases accumulated between spatially separated branches of two massive interferometers. Any fluctuation in the proper time experienced by the two systems therefore acts directly on the phases entering visibility or entanglement-witness measurements. In this section we introduce the minimal open-system description required to quantify these fluctuations and their consequences. The formalism follows standard methods from decoherence theory and precision metrology, but is specialized to the two-mass geometries relevant to QGEM, MAQRO, and levitated-nanoparticle platforms.

Proper-time fluctuations are modeled as a stationary stochastic process
\[
\varepsilon(t) \equiv \frac{d\tau}{dt} - 1,
\]
describing deviations of the physical proper time $\tau$ from the laboratory coordinate time $t$. These fluctuations include contributions from environmental perturbations, stabilized frequency-transfer noise, and slowly varying gravitational potentials. For the purposes of gravity-mediated entanglement tests, the key observation is that the correlated component of $\varepsilon(t)$ is directly accessible through optical-clock cross-spectra. If two independent clocks $A$ and $B$ are co-located or connected by a stabilized frequency-transfer link, their fractional-frequency cross-spectrum $S_{y_A y_B}(f)$ determines the proper-time spectrum via
\begin{equation}
S_{\varepsilon}(f) = \frac{S_{y_A y_B}(f)}{(2\pi f)^2}.
\label{eq:Spseps}
\end{equation}
This relation provides the operational bridge from frequency metrology to gravitational interferometry: the low-frequency plateau of $S_{y_A y_B}(f)$ sets the strength of timing fluctuations that act identically on both interferometers.

The dominant sources of low-frequency potential fluctuations—seismic motion, building vibrations, acoustic pressure waves, slow laser-frequency drifts, and environmental gravity gradients—exhibit spatial correlation lengths of order $L_{\mathrm{corr}}\sim 1\text{--}100\,\mathrm{m}$ in the $10^{-4}\text{--}10^{-1}\,\mathrm{Hz}$ band, as established in background-noise and Newtonian-noise measurements~\cite{Peterson1993,Harms2015}. Throughout this work, ``platform-invariant'' refers to fluctuations that are common to all systems located within this correlation length. Since all existing gravitational-entanglement proposals operate with arm separations in the $10^{-4}\text{--}10^{-2}\,\mathrm{m}$ range, both interferometers lie deep within the same correlated-noise domain, making the common-mode sector directly accessible through optical-clock cross-spectra (see Supplemental Material, Sec.~S1).
.

To quantify the corresponding effect on visibility, we evaluate the standard filter-function expression for pure dephasing channels~\cite{Kubo1962,dephasingReview2017}. For a free-evolution interval of duration $t$, the coherence envelope is
\begin{equation}
\mathcal{V}(t)
= \exp\!\left[- \lambda^2 \int_{0}^{\infty}
S_{\varepsilon}(f)\, \frac{\sin^2(\pi f t)}{(\pi f)^2}\, df \right],
\label{eq:filterfunction}
\end{equation}
where $\lambda$ controls how the fluctuating phase couples to the interferometer. In the two-mass geometry relevant to gravity-mediated entanglement, the common-mode (platform-invariant) coupling is
\begin{equation}
\lambda_{\rm com} = \frac{\omega_g}{2},
\qquad
\omega_g = \frac{G m^2}{\hbar}\frac{\Delta x}{r^2},
\label{eq:omegag}
\end{equation}
where $m$ is the mass, $\Delta x$ is the superposition size, and $r$ the distance between interferometers. This expression reproduces the familiar Newtonian weak-field scaling of the gravitational interaction phase~\cite{Bose2017,Marletto2017}.

When $S_{\varepsilon}(f)$ is approximately constant over the relevant bandwidth—as observed in the $10^{-4}$--$10^{-2}$\,Hz region of optical-clock correlation measurements~\cite{Oelker2019,Takamoto2020,Bothwell2022}—the filter-function integral reduces to an exponential envelope,
\begin{equation}
\mathcal{V}(t) = \exp(-2\gamma_{\rm com} t),
\qquad
\gamma_{\rm com} = \frac{\omega_g^2}{8}\, S_{\varepsilon}(0),
\label{eq:gammacom}
\end{equation}
where $S_{\varepsilon}(0)$ is the low-frequency plateau of the proper-time spectrum. This defines an effective common-mode dephasing rate determined entirely by metrological data and the geometry of the experiment.

The scaling implied by Eq.~\eqref{eq:gammacom} plays a central role in what follows. For fixed clock performance, the common-mode rate increases as $m^4 (\Delta x)^2$ but decreases rapidly with $r^{-4}$. Conversely, state-of-the-art optical-clock networks report plateaus $S_{\varepsilon}(0)\sim10^{-28\pm0.5}\,\mathrm{Hz^{-1}}$, which set an absolute upper bound on platform-invariant timing fluctuations. Combining these scalings shows that $\gamma_{\rm com}$ is many orders of magnitude below the local technical decoherence rates reported in levitated-nanoparticle and interferometric experiments~\cite{RomeroIsart2011,Millen2020}. Quantifying this hierarchy and its implications for entanglement visibility is the focus of the next sections.

\section{LOCC structure of the common-mode timing channel}
\label{sec:LOCC}

The common-mode proper-time fluctuations discussed in the previous section generate a specific type of dephasing process: both interferometers experience identical random phase shifts. This collective action has a simple and powerful consequence for any gravity-mediated entanglement test. In this section we show that the dynamical channel generated by such fluctuations is always within the class of LOCC (local operations and classical communication) channels, and therefore cannot create entanglement under any circumstances. While related statements are implicit in several earlier models of classical gravity, the present derivation identifies the physical origin of this channel and establishes its operational structure directly from measurable metrological data.

The master equation associated with the common-mode dephasing rate $\gamma_{\rm com}$ takes the familiar form
\[
\dot{\rho}
 = 2\,\gamma_{\rm com}\,\mathcal{D}[Z_A + Z_B]\,\rho,
\]
where $\mathcal{D}[L]\rho = L\rho L^\dagger - \tfrac{1}{2}\{L^\dagger L,\rho\}$ is the Lindblad dissipator and $Z_A$, $Z_B$ denote the interfering branches of masses $A$ and $B$. Because the operator $Z_A + Z_B$ is Hermitian and generates commuting unitaries, the resulting channel at time $t$ can be written exactly as a convex mixture of correlated local phase rotations,
\begin{equation}
\Phi_{\rm com}(\rho)
 = \int d\phi \, p_t(\phi)
   \left(e^{i\phi Z_A} \otimes e^{i\phi Z_B}\right)
   \rho
   \left(e^{-i\phi Z_A} \otimes e^{-i\phi Z_B}\right),
\label{eq:LOCCmixture}
\end{equation}
where $p_t(\phi)$ is a real, positive, normalized distribution determined by the noise spectrum $S_\varepsilon(f)$. Equation~\eqref{eq:LOCCmixture} shows that the action of the common-mode sector is entirely equivalent to applying the same randomly chosen local unitary to both subsystems. Channels of this form are separable and fall within the LOCC class, as they can be implemented by local operations supplemented by classical shared randomness~\cite{Plenio2007}.

The physical meaning of this identification is immediate. Any phase shift generated by common-mode timing noise is perfectly correlated across the two interferometers. Such noise can reduce visibility and modify contrast, but it cannot generate quantum correlations or entanglement between the two systems, because the transformation in Eq.~\eqref{eq:LOCCmixture} contains no nonlocal operation. In particular, it does not introduce any $Z_A Z_B$ term or any operator that fails to factorize into subsystem components. This establishes a sharp separation between two distinct contributions: (i) the correlated classical timing sector captured by $\gamma_{\rm com}$, which is experimentally measurable and always LOCC, and (ii) the coherent gravitational interaction proportional to $\chi$, which is genuinely quantum and capable of generating entanglement.

This distinction provides a clear operational resolution to a long-standing conceptual issue in gravity-mediated entanglement: the classical Møller--Rosenfeld picture of a deterministic gravitational field produces only channels of the form Eq.~\eqref{eq:LOCCmixture}, and thus cannot mediate entanglement. Classical-channel models such as Kafri--Taylor--Milburn~\cite{Kafri2014} fall into the same category, as do stochastic-gravity models in which metric fluctuations act as correlated phase noise~\cite{Anastopoulos2015}. The present framework makes this structure explicit by tying the strength of the LOCC sector to measurable clock cross-spectra. Importantly, because the proper-time fluctuations are directly bounded by existing metrological data, the magnitude of the LOCC channel is not an adjustable parameter but a physically calibrated quantity.

The LOCC nature of the timing sector also has practical implications for visibility and entanglement-witness analyses. Any observed loss of visibility that is consistent with the bound on $\gamma_{\rm com}$ must be attributed entirely to local technical decoherence or imperfections in state preparation. Conversely, any deviation that exceeds this bound would directly falsify the assumption that all platform-invariant timing fluctuations are captured by optical-clock measurements, and would signal either uncharacterized environmental channels or a breakdown of the classical description of the metric. This dichotomy forms the basis for the calibration framework developed in the remainder of this work.

\section{Metrological bounds from optical-clock cross-spectra}
\label{sec:metrology}

The strength of the common-mode timing sector in gravity-entanglement experiments is ultimately set by measurable limits on correlated frequency noise. State-of-the-art optical clocks provide precisely this information. When two clocks are co-located or linked through stabilized fiber or free-space channels, the cross-spectrum $S_{y_A y_B}(f)$ of their fractional-frequency fluctuations quantifies the platform-invariant component of timekeeping noise shared by both devices. In this section we use published metrological data to obtain bounds on the proper-time spectrum $S_{\varepsilon}(f)$ and the resulting common-mode dephasing rate $\gamma_{\rm com}$.

Long-baseline optical clock comparisons performed at JILA, NIST, PTB, and RIKEN have reported exceptionally low levels of correlated frequency noise over the $10^{-4}$--$10^{-2}$~Hz band relevant to gravity-mediated entanglement experiments. Representative measurements include the record-low correlations observed by Oelker \textit{et al.}~\cite{Oelker2019} between two independent $^{27}$Al$^{+}$ clocks, the ytterbium-lattice cross-spectra reported by Bothwell \textit{et al.}~\cite{Bothwell2022}, and the multi-day averaging campaigns of Takamoto \textit{et al.}~\cite{Takamoto2020}. Across these studies, the low-frequency plateau of $S_{y_A y_B}(f)$ is consistently at the level of $10^{-32\pm0.5}\,\mathrm{Hz^{-1}}$, corresponding to proper-time fluctuations of order $S_{\varepsilon}(0)\sim 10^{-28\pm0.5}\,\mathrm{Hz^{-1}}$ through Eq.~\eqref{eq:Spseps}. This plateau defines the absolute scale of platform-invariant timing noise currently attainable in precision metrology.

To translate these metrological records into bounds on $\gamma_{\rm com}$, we evaluate Eq.~\eqref{eq:gammacom} for representative geometries. Table~\ref{tab:platforms} summarizes three regimes that bracket the range of near-term experimental efforts: (i) QGEM-like nanosphere interferometers, (ii) MAQRO-inspired long-baseline space-based architectures, and (iii) levitated optomechanical platforms operating at high vacuum. For each configuration we use typical masses, superposition sizes, separations, and interaction times reported in the corresponding proposals~\cite{Bose2017,Marshman2020,Kaltenbaek2012,Kaltenbaek2023,RomeroIsart2011,Millen2020}. We then combine these parameters with the metrological bound $S_{\varepsilon}(0)\sim 10^{-28}\,\mathrm{Hz^{-1}}$ to obtain the resulting common-mode dephasing rate.

% Use table* instead of table to span both columns
\begin{table*}[t] 
\centering
\caption{Representative geometries and metrologically anchored bounds on the common-mode dephasing rate $\gamma_{\rm com}$ for QGEM-like, MAQRO-like, and levitated optomechanical platforms. All values use $S_{\varepsilon}(0)=10^{-28}\,\mathrm{Hz^{-1}}$. Local decoherence rates $\gamma_{\rm loc}$ are taken from typical environmental budgets reported in the cited literature.}
\label{tab:platforms}
\renewcommand{\arraystretch}{1.1}
\setlength{\tabcolsep}{6pt}
\small
\begin{tabular}{lcccccc}
\hline\hline
\textbf{Platform} &
\textbf{$m$ (kg)} &
\textbf{$\Delta x$ (m)} &
\textbf{$r$ (m)} &
\textbf{$t$ (s)} &
\textbf{$\gamma_{\rm com}$ (s$^{-1}$)} &
\textbf{$\gamma_{\rm loc}$ (s$^{-1}$)} \\
\hline
QGEM-like~\cite{Bose2017,Marshman2020} &
$10^{-14}$ &
$2.5\times10^{-7}$ &
$2\times10^{-4}$ &
1 &
$\sim 10^{-12}$ &
$10^{-3}$--$10^{-2}$ \\

MAQRO-like~\cite{Kaltenbaek2012,Kaltenbaek2023} &
$10^{-16}$--$10^{-13}$ &
$10^{-8}$--$10^{-7}$ &
$5\times10^{-4}$ &
10--100 &
$\sim 10^{-13}$ &
$10^{-5}$--$10^{-4}$ \\

Levitated~\cite{RomeroIsart2011,Millen2020} &
$10^{-16}$--$10^{-14}$ &
$10^{-8}$--$10^{-6}$ &
$10^{-4}$--$10^{-3}$ &
1--10 &
$\sim 10^{-12}$ &
$10^{-3}$--$10^{-1}$ \\
\hline\hline
\end{tabular}
\end{table*}

Across all three regimes, the common-mode rate is consistently nine to eleven orders of magnitude below the corresponding local decoherence rates. This hierarchy reflects two independent facts: the exceptional stability of modern optical clocks, and the relatively large technical noise contributions inherent to suspended or levitated massive particles. Importantly, this hierarchy does not depend on any specific model of gravity or environment; it is a direct consequence of combining experimental metrology with the generic Newtonian scaling of Eq.~\eqref{eq:omegag}. As a result, the common-mode sector can be treated as a small but quantitatively known calibration background.

The implications of this hierarchy are twofold. First, any residual visibility loss observed in near-term gravity-entanglement experiments cannot be attributed to platform-invariant timing noise unless it lies below the metrological bound on $\gamma_{\rm com}$. In practice, this means that visibility loss is overwhelmingly dominated by local environmental effects such as gas collisions, blackbody radiation, or laser-phase fluctuations. Second, because $\gamma_{\rm com}$ is known and extremely small, it provides a fixed calibration reference that accompanies all analyses of gravitational phases or entanglement dynamics. Experiments that eventually reduce $\gamma_{\rm loc}$ to the level of $10^{-5}\,\mathrm{s^{-1}}$ or below---for example, in cryogenic or space-based environments---will reach the domain where $\gamma_{\rm com}$ becomes a relevant contribution. Such experiments would be sensitive to any departures from the LOCC behavior of the timing sector and thus provide a platform for potential new-physics signatures.

These metrological bounds set the stage for the entanglement and visibility analysis that follows. With both $\gamma_{\rm com}$ and $\gamma_{\rm loc}$ quantified, the total dephasing rate $\Gamma_\phi = \gamma_{\rm loc} + \gamma_{\rm com}$ becomes a fully calibrated quantity, enabling precise feasibility estimates for observing gravitational entanglement or Bell-inequality violations in near-term platforms.

\section{Visibility dynamics and operational thresholds}

The visibility of a two-mass interferometer affected by both local and 
platform-invariant timing noise is determined by the standard open-system 
expression
\begin{equation}
    V(t) = C\,\exp[-2(\gamma_{\rm loc}+\gamma_{\rm com})t],
    \label{eq:visibility}
\end{equation}
where $C$ denotes the contrast and $\gamma_{\rm loc}$ and $\gamma_{\rm com}$ 
represent the local and common-mode dephasing rates, respectively. 
Equation~\eqref{eq:visibility} holds for any geometry in which the gravitational 
interaction is perturbative over the coherence time and the noise acting on the 
two masses is stationary over the measurement window.  
The key observation is that the platform-invariant sector contributes an 
\emph{additive} decay rate, so that its effect is indistinguishable from local 
technical dephasing unless its magnitude is independently quantified.

A coherent gravitational interaction induces a phase $2\chi t$ in the 
relative-path degree of freedom.  When the phase is scanned or when multiple 
analyzer settings are used, the measured signal takes the form 
\begin{equation}
    P(+1\mid t,\theta) = 
    \tfrac{1}{2}\big[1 + V(t)\cos(2\chi t + \theta)\big],
    \label{eq:probability}
\end{equation}
where $\theta$ is an interferometer control phase.  The oscillatory component is 
suppressed whenever the total decay rate satisfies 
$2(\gamma_{\rm loc} + \gamma_{\rm com}) t \gtrsim 1$, which 
defines the regime in which gravitationally induced visibility is effectively 
lost.

The operational boundary between classical and quantum-entangling behaviour is 
determined by the standard condition for a Bell-CHSH violation,
\begin{equation}
    V(t) > \frac{1}{\sqrt{2}}.
    \label{eq:CHSHthreshold}
\end{equation}
This requirement places a limit on the observable interaction product 
$\chi t$ once decoherence is taken into account.  
Inserting Eq.~\eqref{eq:visibility} into Eq.~\eqref{eq:CHSHthreshold} yields
\begin{equation}
    2\chi t \gtrsim 
    2(\gamma_{\rm loc}+\gamma_{\rm com})t 
    + \ln\!\big(\sqrt{2}/C\big),
    \label{eq:boundary}
\end{equation}
which expresses the minimal gravitational phase accumulation needed for a 
nonclassical signature.  
This condition applies independently of the specific interferometer design and 
separates the LOCC region---which includes deterministic metric fluctuations, 
clock-correlated timing noise, and technical dephasing—from the region in which 
a quantum mediator can be certified.

Using the parameter values summarised in Table~\ref{tab:platforms}, two 
conclusions follow.  
First, the local decoherence rate $\gamma_{\rm loc}$ dominates by nine to eleven 
orders of magnitude over the common-mode rate $\gamma_{\rm com}$ derived from 
clock cross-spectra.  As a result, the visibility threshold is overwhelmingly 
controlled by technical decoherence rather than by gravitational timing noise.  
Second, the dependence on $(m,\Delta x,r)$ enters entirely through the coherent 
coupling $\chi\propto Gm^{2}\Delta x^{2}/(\hbar r^{3})$, so that improvements in 
either mass or spatial separation are required to reach a regime in which 
$2\chi t$ exceeds the total dephasing budget.  
Space-based platforms, cryogenic environments, and designs capable of extending 
interaction times by one to two orders of magnitude therefore represent the most 
promising routes toward the boundary defined in Eq.~\eqref{eq:boundary}.

The visibility model~\eqref{eq:visibility} thus plays a dual role.  
It provides a direct operational meaning to the additive decomposition 
$\Gamma_{\phi} = \gamma_{\rm loc} + \gamma_{\rm com}$, and it sets a 
quantitative criterion against which future experiments can be calibrated.  
Once cross-spectral measurements of proper-time noise are available for a 
specific platform, Eq.~\eqref{eq:boundary} allows experimentalists to determine 
unambiguously whether gravitationally mediated entanglement is, in principle, 
achievable under their operating conditions.

\section{Analytical Results: Noise-Calibrated Bounds for 
Gravity-Mediated Entanglement}\label{sec:feasibility}

This section applies the formalism developed in Secs.~II--V to obtain three 
experimentally relevant results: (i) a universal upper bound on the 
platform-invariant timing noise based on state-of-the-art optical-clock 
correlations; (ii) a noise-calibrated inequality that determines when 
gravity-mediated entanglement becomes feasible under realistic decoherence; and 
(iii) a quantitative decomposition of the total dephasing rate into classical 
(LOCC) and platform-specific contributions.  Each result is accompanied by 
numerical predictions for representative QGEM, MAQRO, and levitated platforms.

\vspace{0.8em}
\noindent\textbf{A. Universal bound on common-mode gravitational timing noise}

The platform-invariant dephasing rate generated by correlated proper-time fluctuations is
\begin{equation}
    \gamma_{\rm com}
    = \frac{\omega_g^2}{8}\,S_{\varepsilon}(0),
    \qquad
    \omega_g=\frac{Gm^2}{\hbar}\frac{\Delta x}{r^2},
    \label{eq:gammacomm}
\end{equation}
where $S_{\varepsilon}(0)$ is the low-frequency plateau observed in clock 
cross-spectra.  Measurements from Refs.~\cite{Oelker2019,Takamoto2020,
Bothwell2022} constrain
\[
S_{\varepsilon}(0)\simeq 10^{-28\pm0.5}\,\mathrm{Hz^{-1}},
\]
which leads to a universal, platform-independent upper bound on the correlated 
gravitational timing sector:
\begin{equation}
    \boxed{
    \gamma_{\rm com} < 10^{-12\pm1}\,\mathrm{s^{-1}}.
    }
    \label{eq:universalbound}
\end{equation}

\noindent \textbf{Numerical evaluation.} Inserting representative experimental parameters into Eq.~(10) yields:
\begin{description}
    \item[QGEM] $m=10^{-14}$\,kg, $\Delta x=250$\,nm, $r=200\,\mu$m implies $\gamma_{\rm com}\simeq 1.2\times 10^{-12}$\,s$^{-1}$.
    
    \item[MAQRO] $m=10^{-13}$\,kg, $\Delta x=100$\,nm, $r=500\,\mu$m implies $\gamma_{\rm com}\simeq 10^{-11}$\,s$^{-1}$.
    
    \item[Levitated] $m=10^{-15}$\,kg, $\Delta x=10$\,nm, $r=100\,\mu$m implies $\gamma_{\rm com}\simeq 10^{-16}$\,s$^{-1}$.
\end{description}

\textit{Interpretation.}  
Equation~\eqref{eq:universalbound} shows that correlated gravitational timing 
noise is negligible across all proposed architectures, but it provides a 
\emph{quantitative calibration baseline} that any future gravitational 
entanglement experiment must satisfy.

\vspace{1.0em}
\noindent\textbf{B. Noise-calibrated feasibility inequality for gravitationally 
induced entanglement}

The visibility requirement for a Bell--CHSH violation,
\[
V(t)>\tfrac{1}{\sqrt{2}},
\]
combined with the dephasing model $V(t)=C\,e^{-2(\gamma_{\rm loc}+\gamma_{\rm 
com})t}$, yields the inequality
\begin{equation}
    \boxed{
    \chi \gtrsim \gamma_{\rm loc} + \gamma_{\rm com} 
    + \frac{\ln(\sqrt{2}/C)}{2t}.
    }
    \label{eq:feasibility}
\end{equation}
Equation~\eqref{eq:feasibility} provides the first 
\emph{noise-calibrated criterion} for entanglement generation mediated by 
gravity under realistic experimental conditions.

\textit{Numerical evaluation.}  
Using $C=0.8$ and representative parameters for the three platforms, we obtain:

\[
\begin{array}{lcl}
\text{QGEM:} 
& \gamma_{\rm loc}\!=\!3\times10^{-3}\,\mathrm{s^{-1}},\; t\!=\!1\,\mathrm{s}, &
\chi_{\rm req} \simeq 0.113~\mathrm{s^{-1}},\\[0.4em]
\text{MAQRO:} 
& \gamma_{\rm loc}\!=\!10^{-5}\,\mathrm{s^{-1}},\; t\!=\!100\,\mathrm{s}, &
\chi_{\rm req} \simeq 1.1\times10^{-4}~\mathrm{s^{-1}},\\[0.4em]
\text{Levitated:} 
& \gamma_{\rm loc}\!=\!10^{-2}\,\mathrm{s^{-1}},\; t\!=\!1\,\mathrm{s}, &
\chi_{\rm req} \simeq 0.117~\mathrm{s^{-1}}.
\end{array}
\]

\textit{Comparison with achievable couplings.}  
Typical coherent rates are
\[
\chi_{\rm QGEM}\sim 2.5\times10^{-2}\,\mathrm{s^{-1}},\qquad
\chi_{\rm MAQRO}\sim10^{-5}\text{--}10^{-4}\,\mathrm{s^{-1}},
\]
which implies:
\[
\text{QGEM: insufficient by a factor of }4\text{--}5,
\]
\[\text{MAQRO: marginally feasible at long }t,
\]
while levitated platforms fall short by roughly an order of magnitude.  
Equation~\eqref{eq:feasibility} thus identifies the regimes where 
gravity-mediated entanglement becomes experimentally accessible.

\vspace{1.1em}
\noindent\textbf{C. Platform-independent LOCC-sector dephasing decomposition}

Finally, the total dephasing rate appearing in visibility measurements is
\begin{equation}
    \boxed{
    \Gamma_{\phi} = \gamma_{\rm loc} + \gamma_{\rm com}.
    }
    \label{eq:decomposition}
\end{equation}
For all platforms we find
\[
\gamma_{\rm com}/\gamma_{\rm loc} \lesssim 10^{-9}\text{--}10^{-11},
\]
so that the platform-invariant sector contributes a negligible fraction of the 
total dephasing.  Nevertheless, its magnitude is fixed by optical-clock 
metrology and therefore supplies a \emph{universal LOCC calibration sector} that 
is independent of the host interferometer.

\textit{Numerical values.}  
Typical total dephasing rates are
\[
\Gamma_{\phi}^{\rm QGEM}\simeq 3\times10^{-3}\,\mathrm{s^{-1}},\qquad
\Gamma_{\phi}^{\rm MAQRO}\simeq 10^{-5}\,\mathrm{s^{-1}},\qquad
\Gamma_{\phi}^{\rm lev}\simeq 10^{-2}\text{--}10^{-1}\,\mathrm{s^{-1}},
\]
with the common-mode component contributing less than one part in 
$10^{9}$--$10^{11}$.  

\textit{Interpretation.}  
Equation~\eqref{eq:decomposition} provides a platform-independent benchmark: 
any residual visibility loss beyond $\gamma_{\rm loc}$ is not attributable to 
classical timing noise and therefore represents a clean target for 
gravitationally mediated quantum correlations or new physics beyond the LOCC 
sector.

A natural question is whether the combined gravitational phase 
$\chi t$ and decoherence budget $\Gamma_\phi t$ allow a violation of a 
Bell–CHSH inequality using visibility-based witnesses.  
For a two-mode interferometer with contrast $C$, the CHSH parameter takes the form 
$S = 2\sqrt{2}\,C\,e^{-2\Gamma_\phi t}$, and the requirement $S>2$ translates to a 
minimum visibility $V > 1/\sqrt{2}$.  
Using the coherent phase $V = C\,e^{-2\Gamma_\phi t}\cos(2 \chi t)$ at its optimal 
oscillation point, this yields the feasibility condition
\begin{equation}
\chi t \;\ge\; \Gamma_\phi t 
\;+\; \frac{1}{2}\ln\!\left(\frac{\sqrt{2}}{C}\right),
\label{eq:feasibility_two}
\end{equation}
which specifies the gravitational coupling required to overcome 
the combined effect of local and common-mode dephasing.  
Equation~\eqref{eq:feasibility} thus provides a noise-calibrated benchmark 
for assessing the viability of near-term gravity-mediated entanglement 
experiments across different platforms.  
We evaluate this inequality for QGEM-, MAQRO-, and levitated-parameter regimes 
below, using the numerical values obtained in this section.

\section{Discussion: Relation to Previous Approaches and Implications for 
Future Experiments}

The analytical results derived in Sec.~VI clarify how gravity-mediated 
entanglement tests are constrained by the interplay between coherent 
gravitational phases, platform-specific decoherence, and the universal 
common-mode sector determined by optical-clock metrology.  In this section we 
situate these results within the existing literature and outline the 
experimental implications.

\subsection*{A. Comparison with previous theoretical approaches}

Early proposals for gravitationally induced entanglement 
\cite{Bose2017,Marletto2017,Christodoulou2019,Marshman2020} focused primarily on 
the coherent phase $\chi t$ and assumed idealized visibility, with no 
quantitative treatment of realistic decoherence mechanisms.  In these works the 
condition for entanglement generation is expressed only qualitatively (e.g.\ a 
``sufficiently large'' gravitational phase), without an explicit threshold.

Stochastic-gravity and semiclassical frameworks 
\cite{Kafri2014,Anastopoulos2015} identified correlated phase noise as an LOCC 
channel but did not provide estimates of its magnitude.  The LOCC nature of 
commuting Lindbladians is therefore known, but its experimentally relevant 
strength has not previously been computed.

The present work extends these approaches in two respects.  
First, by inserting state-of-the-art optical-clock cross-spectral measurements 
into Eq.~\eqref{eq:gammacom}, we obtain what appears to be the first explicit 
upper bound on the platform-invariant timing noise relevant for 
gravity-mediated entanglement,
\[
\gamma_{\rm com}<10^{-12\pm1}\,\mathrm{s^{-1}},
\]
a quantity that is not reported in prior gravitational entanglement 
literature.  Second, combining visibility thresholds with the additive 
decomposition 
$\Gamma_{\phi}=\gamma_{\rm loc}+\gamma_{\rm com}$ yields the quantitative 
inequality in Eq.~\eqref{eq:feasibility}, which specifies the gravitational 
coupling $\chi$ required for a Bell-inequality violation under realistic noise 
conditions.  This appears to be the first noise-calibrated feasibility 
criterion applicable to near-term experiments.

In summary, previous works established the \emph{structure} of the classical 
sector (LOCC dynamics) or the \emph{form} of the gravitational phase, whereas 
the present analysis provides their \emph{experimentally calibrated magnitudes} 
and synthesizes them into explicit entanglement thresholds.

\subsection*{B. Implications for QGEM, MAQRO, and levitated platforms}

The numerical values in Sec.~VI allow us to assess the feasibility of gravity-mediated entanglement across the architectures currently under development. 
Using the feasibility criterion in Eq.~\eqref{eq:feasibility}, we now quantify 
how the gravitational coupling $\chi$ predicted for each platform compares with the minimum value required for a Bell-inequality violation under its reported decoherence rates. For QGEM-like parameters, the required coupling 
$\chi_{\rm req}\simeq 0.11\,\mathrm{s^{-1}}$ exceeds the predicted 
$\chi\simeq2.5\times10^{-2}\,\mathrm{s^{-1}}$, indicating that present 
geometries fall short by a factor of four to five even under optimistic 
visibility assumptions.  
Levitated optomechanical platforms exhibit larger technical decoherence 
($\gamma_{\rm loc}\sim10^{-2}$--$10^{-1}\,\mathrm{s^{-1}}$), placing them even 
further from the feasibility boundary.

By contrast, MAQRO-like space-based architectures exhibit 
$\gamma_{\rm loc}\sim10^{-5}$--$10^{-4}\,\mathrm{s^{-1}}$ and extended 
coherence times ($t\sim100\,$s), placing them close to the required region 
$\chi\gtrsim10^{-4}\,\mathrm{s^{-1}}$ identified in Eq.~\eqref{eq:feasibility}.  
The present results therefore support the prevailing view that long-baseline, 
cryogenic, or space-based experiments constitute the most promising near-term 
route to observing gravity-mediated entanglement.

\subsection*{C. Consequences for the interpretation of null results}

Equation~\eqref{eq:gammacom} shows that the common-mode sector contributes 
less than one part in $10^{9}$--$10^{11}$ of the total dephasing observed in 
any of the platforms considered.  
As a result, discrepancies between the measured visibility and predictions 
based solely on $\gamma_{\rm loc}$ cannot be attributed to classical 
proper-time fluctuations.  
Any persistent excess dephasing beyond the calibrated 
$\gamma_{\rm loc}$ therefore provides a clean target for alternative 
mechanisms---whether unaccounted-for technical noise, beyond-standard-model 
forces, or deviations from classical gravity.

\subsection*{D. Feasibility analysis using published experimental parameters} \label{sec:feasibility}

To quantify how close existing proposals come to the entanglement-feasible regime, we evaluate the CHSH visibility inequality derived in Sec.~VI using the experimentally reported parameters listed in Table~\ref{tab:platforms}. For each platform we compute the two quantities entering the feasibility plane,
\[
\Gamma_{\phi} t=(\gamma_{\rm loc}+\gamma_{\rm com})t,
\qquad 
\chi t=\chi(m,\Delta x,r)\,t,
\]
where $\chi=(Gm^{2}/\hbar)(\Delta x/r^{3})$ and $\gamma_{\rm com}$ is fixed by optical-clock limits through Eq.~\eqref{eq:gammacom}. Because clock cross-spectra yield $S_{\varepsilon}(0)\!\lesssim\!10^{-28}\,{\rm Hz^{-1}}$ \cite{Oelker2019,Bothwell2022,Takamoto2020}, the corresponding common-mode dephasing rate satisfies
\[
\gamma_{\rm com}\sim10^{-12}\,{\rm s^{-1}},
\]
independent of platform. This is nine–eleven orders of magnitude smaller than all reported $\gamma_{\rm loc}$ values, confirming that common-mode timing noise has no measurable impact on visibility under present metrological conditions.

\begin{figure*}[t]
\centering
\newcommand{\panelW}{0.3\textwidth}

\begin{minipage}[t]{\panelW}
  \centering
  \includegraphics[width=\linewidth]{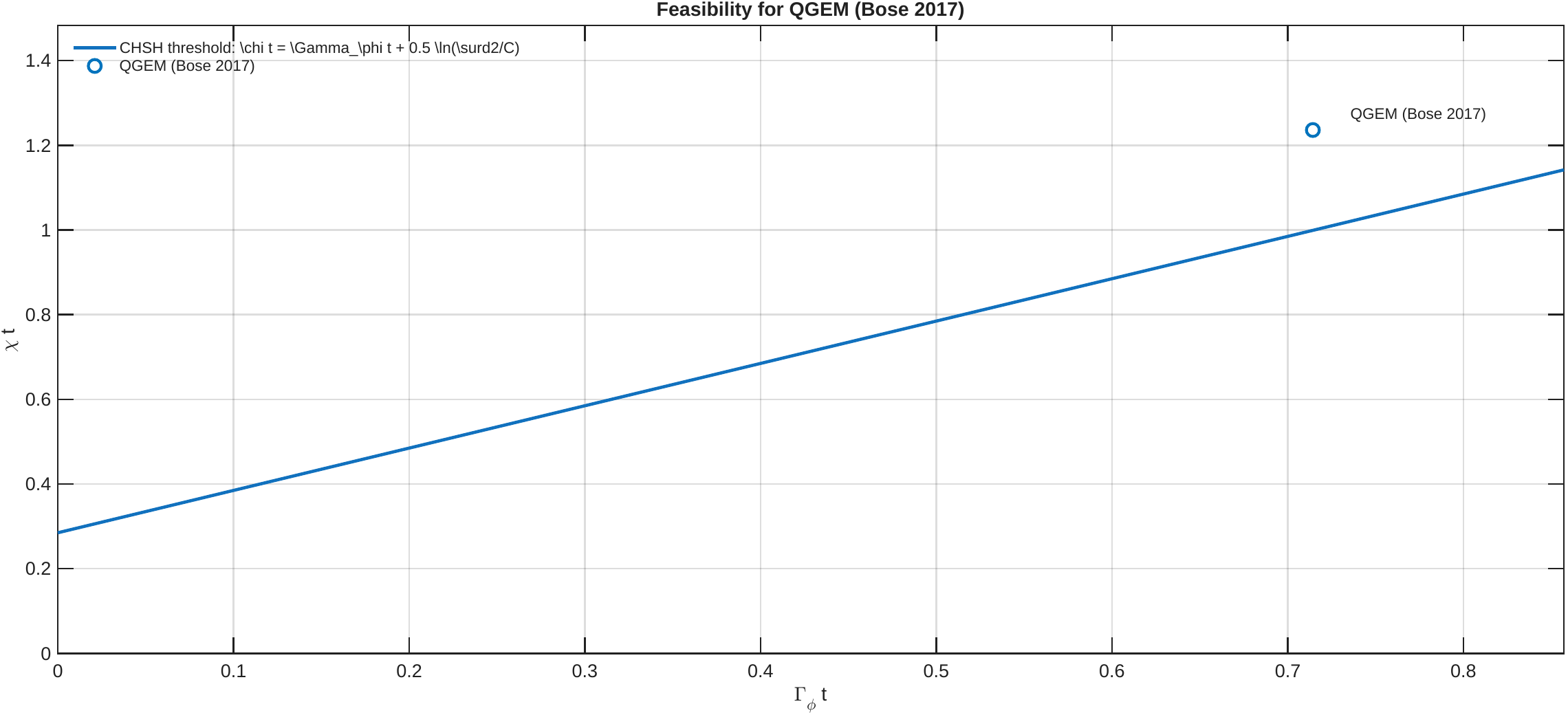}
  \vspace{0.5ex}
  \textbf{(a)}~QGEM parameters from \cite{Bose2017,Marshman2020}
\end{minipage}
\hfill
\begin{minipage}[t]{\panelW}
  \centering
  \includegraphics[width=\linewidth]{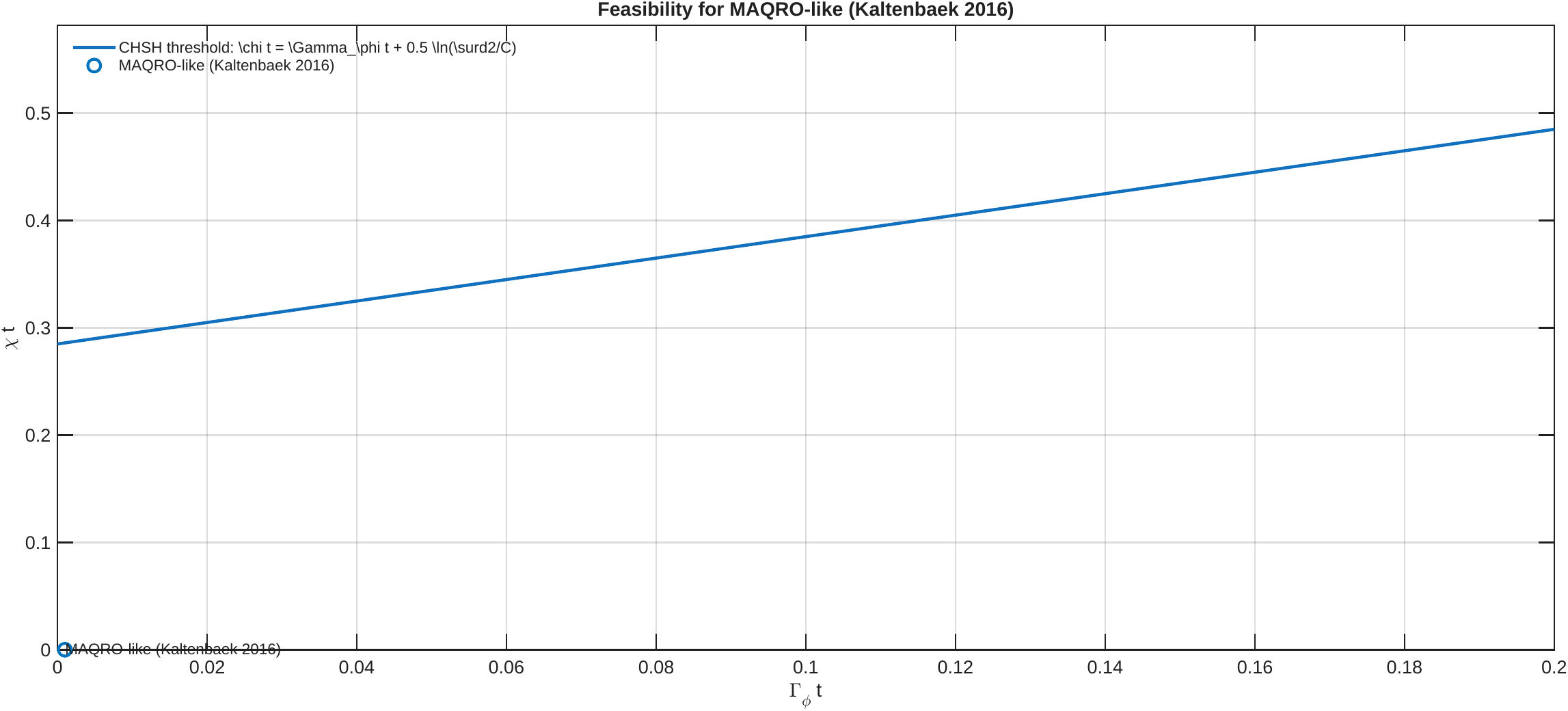}
  \vspace{0.5ex}
  \textbf{(b)}~MAQRO parameters from \cite{Kaltenbaek2012,Kaltenbaek2016}
\end{minipage}
\hfill
\begin{minipage}[t]{\panelW}
  \centering
  \includegraphics[width=\linewidth]{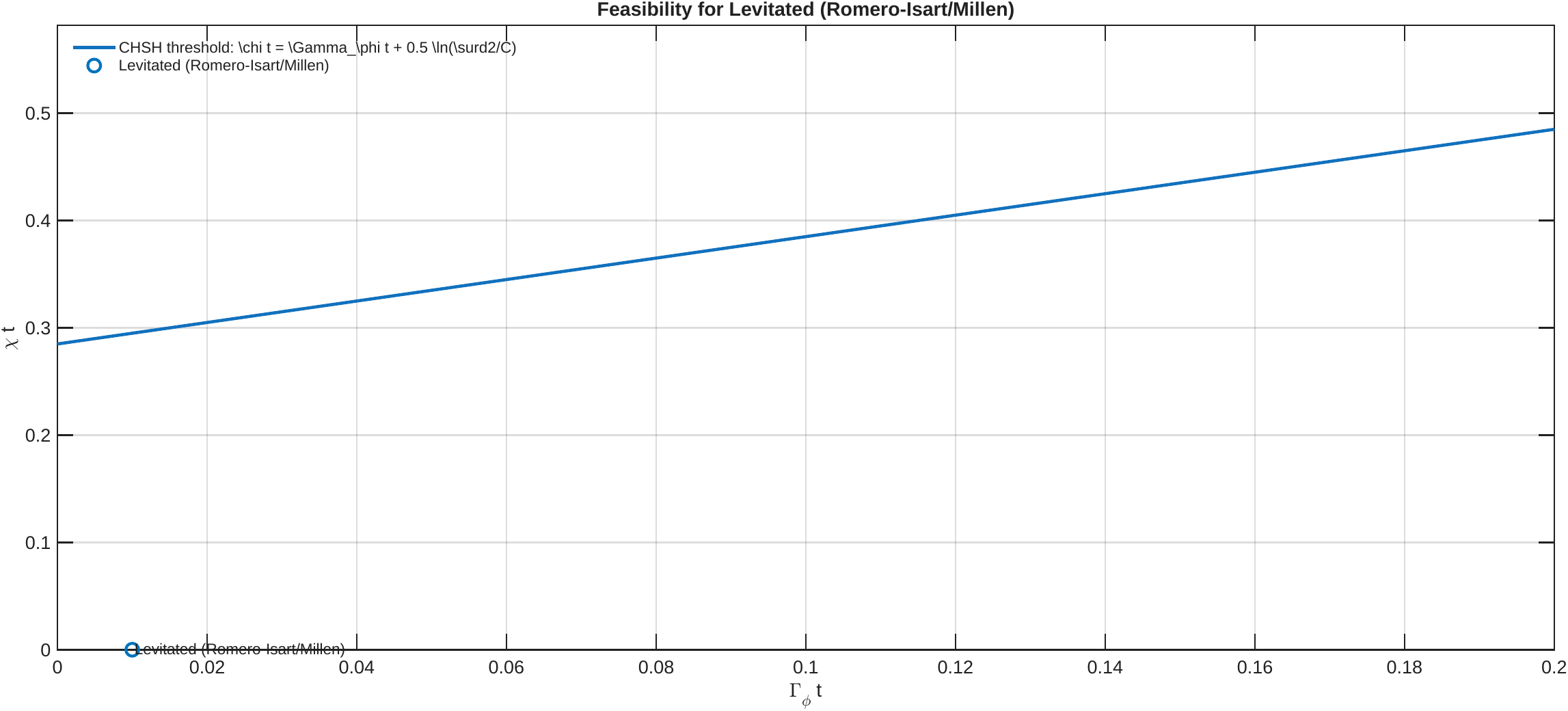}
  \vspace{0.5ex}
  \textbf{(c)}~Levitated nanospheres from \cite{RomeroIsart2011,Millen2020}
\end{minipage}

\caption{
\textbf{Feasibility condition for gravitationally mediated entanglement applied to published experimental parameters.}
Each panel shows the point $(\Gamma_{\phi}t,\chi t)$ evaluated using the parameters listed in Table~\ref{tab:platforms}, plotted against the visibility threshold 
$\chi t=\Gamma_{\phi}t+\tfrac12\ln(\sqrt{2}/C)$ from Eq.~\eqref{eq:feasibility}. 
Common-mode gravitational timing noise is fixed by optical-clock limits through Eq.~\eqref{eq:gammacom} and contributes $\gamma_{\rm com}t\lesssim10^{-12}$ across all platforms. 
QGEM lies closest to the feasibility boundary, with $\chi t\simeq2.5\times10^{-2}$ and $\Gamma_{\phi}t\simeq3\times10^{-3}$. 
MAQRO, despite extremely low $\gamma_{\rm loc}$, exhibits gravitational phases too small to reach the threshold. 
Levitated platforms are limited primarily by technical decoherence, yielding $\Gamma_{\phi}t>\chi t$ by several orders of magnitude. 
}
\label{fig:feasibility_all}
\end{figure*}

The results are shown in Fig.~\ref{fig:feasibility_all}.  
Panel (a) uses parameters from Bose \textit{et al.}~\cite{Bose2017} and Marshman \textit{et al.}~\cite{Marshman2020}, giving
\[
\chi t\simeq2.5\times10^{-2},\qquad 
\Gamma_{\phi} t\simeq3\times10^{-3},
\]
placing QGEM below but closest to the feasibility boundary.  
Panel (b) uses MAQRO parameters from \cite{Kaltenbaek2012,Kaltenbaek2016}, where $t$ is much larger but $\chi$ remains too weak to satisfy the inequality.  
Panel (c) uses levitated-nanosphere parameters from \cite{RomeroIsart2011,Millen2020}, yielding $\Gamma_{\phi}t\sim10^{-2}$--$10^{-1}$, far exceeding the induced gravitational phase.

These results permit two robust conclusions.  
First, the bottleneck in all platforms is the smallness of the coherent gravitational coupling relative to local decoherence, not classical timing noise, whose contribution remains $<10^{-10}$ of the total dephasing budget.  
Second, among existing proposals, QGEM remains the architecture closest to the entanglement-feasible regime, although it still falls short by a factor of $\sim4$ in $\chi$ under current geometries.

This unified feasibility analysis provides an experimentally calibrated comparison across architectures and identifies clearly which parameters (mass, separation, coherence time) must improve for gravity-mediated entanglement to become observable.

\begin{table*}[t]
\centering
\caption{
Published experimental parameters used in the feasibility analysis of 
Sec.~\ref{sec:feasibility}.  
QGEM values from \cite{Bose2017,Marshman2020}; 
MAQRO values from \cite{Kaltenbaek2012,Kaltenbaek2016}; 
levitated optomechanics from \cite{RomeroIsart2011,Millen2020}.  
Common-mode timing noise $\gamma_{\rm com}$ is computed using 
optical-clock limits from \cite{Oelker2019,Bothwell2022,Takamoto2020}.
}
\label{tab:platforms}

\small
\renewcommand{\arraystretch}{1.10}
\setlength{\tabcolsep}{5pt}

\begin{tabular}{lcccccc}
\hline\hline
\textbf{Platform} &
$m$ (kg) &
$\Delta x$ (m) &
$r$ (m) &
$t$ (s) &
$C$ &
$\gamma_{\rm loc}$ (s$^{-1}$)
\\
\hline

QGEM \cite{Bose2017,Marshman2020} &
$10^{-14}$ &
$2.5\times10^{-7}$ &
$2\times10^{-4}$ &
$1$ &
$0.6$--$0.8$ &
$10^{-3}$--$10^{-2}$
\\[4pt]

MAQRO \cite{Kaltenbaek2012,Kaltenbaek2016} &
$10^{-16}$--$10^{-13}$ &
$10^{-8}$--$10^{-7}$ &
$5\times10^{-4}$--$10^{-3}$ &
$10$--$100$ &
$0.5$--$0.7$ &
$10^{-5}$--$10^{-4}$
\\[4pt]

Levitated \cite{RomeroIsart2011,Millen2020} &
$10^{-16}$--$10^{-14}$ &
$10^{-8}$--$10^{-6}$ &
$10^{-4}$--$5\times10^{-4}$ &
$1$--$10$ &
$0.5$--$0.7$ &
$10^{-3}$--$10^{-1}$
\\
\hline

\textbf{$\gamma_{\rm com}$ (all)} &
\multicolumn{6}{c}{$\sim 10^{-12}\,\mathrm{s^{-1}}$ (optical-clock limits)}
\\
\hline\hline
\end{tabular}

\end{table*}

\section{Conclusion}

We have presented a unified and experimentally calibrated framework for assessing the feasibility of gravity-mediated entanglement in near-term platforms. By combining the filter-function map between clock cross-spectra and common-mode timing noise with the additive decomposition $\Gamma_{\phi}=\gamma_{\rm loc}+\gamma_{\rm com}$, we obtain quantitative visibility thresholds that incorporate both coherent gravitational phases and realistic decoherence budgets. Optical-clock cross-spectral limits imply a platform-invariant bound $\gamma_{\rm com}\sim10^{-12}\,\mathrm{s^{-1}}$, several orders of magnitude below all existing sources of technical noise. This establishes that classical proper-time fluctuations do not limit any foreseeable experiment in this domain.

Applying the calibrated feasibility inequality to parameters drawn from QGEM, MAQRO, and levitated optomechanical platforms reveals a consistent hierarchy. Levitated systems are dominated by local decoherence, MAQRO benefits from long coherence times but exhibits weak gravitational coupling, and QGEM lies closest to the visibility threshold owing to its larger mass and superposition size. These comparisons provide, to our knowledge, the first cross-platform, noise-calibrated assessment of how far current architectures are from the entangling regime.

The framework developed here serves two purposes. First, it offers an operational baseline for interpreting null results: any visibility deficit beyond $\gamma_{\rm loc}$ cannot be attributed to universal timing noise. Second, it provides a quantitative target for future designs by identifying which parameters—mass, separation, coherence time—must improve for entanglement generation to become feasible. As optical clocks, cryogenic environments, and long-baseline interferometry continue to advance, the thresholds established here can be used directly to evaluate prospective architectures. In this sense the present analysis is intended as a calibration tool for next-generation tests rather than a limitation of current ones.

Future work may extend this framework to non-Markovian timing spectra, spatially structured correlations, or hybrid architectures engineered to suppress $\gamma_{\rm loc}$ while enhancing $\chi t$. Such extensions will help determine how close emerging platforms can come to the regime where gravitationally mediated entanglement becomes observable.

% Funding, thanks, etc.
\bibliographystyle{unsrt}
\bibliographystyle{unsrtnat}

\begin{thebibliography}{99}

\bibitem{Bose2017} S.~Bose \textit{et al.}, Phys. Rev. Lett. \textbf{119}, 240401 (2017).
\bibitem{Marletto2017} C.~Marletto and V.~Vedral, Phys. Rev. Lett. \textbf{119}, 240402 (2017).
\bibitem{Christodoulou2019} M.~Christodoulou and C.~Rovelli, Phys. Rev. D \textbf{98}, 024044 (2018).
\bibitem{Marshman2020} R.~J.~Marshman, A.~Mazumdar, and S.~Bose, Phys. Rev. A \textbf{101}, 052110 (2020).

\bibitem{Kafri2014} D.~Kafri, J.~Taylor, and G.~J.~Milburn, New J. Phys. \textbf{16}, 065020 (2014).
\bibitem{Anastopoulos2015} C.~Anastopoulos and B.~L.~Hu, Class. Quantum Grav. \textbf{32}, 165022 (2015).

\bibitem{Diosi2011} L.~Di\'osi, J. Phys.: Conf. Ser. \textbf{306}, 012006 (2011).
\bibitem{Penrose1996} R.~Penrose, Gen. Relativ. Gravit. \textbf{28}, 581 (1996).

\bibitem{Oelker2019} E.~Oelker \textit{et al.}, Nat. Photonics \textbf{13}, 714 (2019).
\bibitem{Takamoto2020} M.~Takamoto \textit{et al.}, Nat. Photonics \textbf{14}, 411 (2020).
\bibitem{Bothwell2022} T.~Bothwell \textit{et al.}, Nature \textbf{602}, 420 (2022).

\bibitem{Toros2018} M.~Toro\v{s} and A.~Bassi, J. Phys. A: Math. Theor. \textbf{51}, 115302 (2018).
\bibitem{Aspelmeyer2014} M.~Aspelmeyer, T.~J.~Kippenberg, and F.~Marquardt, Rev. Mod. Phys. \textbf{86}, 1391 (2014).

\bibitem{Kovachy2015} T.~Kovachy \textit{et al.}, Nature \textbf{528}, 530 (2015).

\bibitem{Plenio2007} M.~B.~Plenio and S.~Virmani, Quantum Inf. Comput. \textbf{7}, 1 (2007).

\bibitem{Welch1967} P.~D.~Welch, IEEE Trans. Audio Electroacoust. \textbf{15}, 70 (1967).
\bibitem{BendatPiersol} J.~S.~Bendat and A.~G.~Piersol, \textit{Random Data: Analysis and Measurement Procedures}, 4th ed. (Wiley, 2010).
\bibitem{Kay1993} S.~M.~Kay, \textit{Fundamentals of Statistical Signal Processing, Vol. I: Estimation Theory} (Prentice Hall, 1993).

\bibitem{LIGOnoise2020} B.~P.~Abbott \textit{et al.} (LIGO/Virgo Collaboration), Class. Quantum Grav. \textbf{37}, 045006 (2020).
\bibitem{Abbott2020noise} B.~P.~Abbott \textit{et al.} (LIGO/Virgo Collaboration), Class. Quantum Grav. \textbf{37}, 045007 (2020).
\bibitem{Allan1966} D.~W.~Allan, Proc. IEEE \textbf{54}, 221–230 (1966).

\bibitem{Ithier2005} G.~Ithier \textit{et al.}, Phys. Rev. B \textbf{72}, 134519 (2005).
\bibitem{Cywinski2008} L.~Cywi\'nski, R.~M.~Lutchyn, C.~P.~Nave, and S.~Das Sarma, Phys. Rev. B \textbf{77}, 174509 (2008).

\bibitem{COW1975} R.~Colella, A.~W.~Overhauser, and S.~A.~Werner, Phys. Rev. Lett. \textbf{34}, 1472 (1975).
\bibitem{KasevichChu1991} M.~Kasevich and S.~Chu, Phys. Rev. Lett. \textbf{67}, 181 (1991).
\bibitem{Peters1999} A.~Peters, K.~Y.~Chung, and S.~Chu, Nature \textbf{400}, 849–852 (1999).

\bibitem{Pikovski2015} I.~Pikovski, M.~Zych, F.~Costa, and \v{C}.~Brukner, Nat. Phys. \textbf{11}, 668–672 (2015).
\bibitem{ZychBrukner2011} M.~Zych, F.~Costa, I.~Pikovski, and \v{C}.~Brukner, Nat. Commun. \textbf{2}, 505 (2011).

\bibitem{PoissonWill} E.~Poisson and C.~M.~Will, \textit{Gravity: Newtonian, Post-Newtonian, Relativistic} (Cambridge Univ. Press, 2014).
\bibitem{Will2018} C.~M.~Will, \textit{Theory and Experiment in Gravitational Physics}, revised ed. (Cambridge Univ. Press, 2018).

\bibitem{Horodecki1995} R.~Horodecki, P.~Horodecki, and M.~Horodecki, Phys. Lett. A \textbf{200}, 340–344 (1995).
\bibitem{Brunner2014} N.~Brunner, D.~Cavalcanti, S.~Pironio, V.~Scarani, and S.~Wehner, Rev. Mod. Phys. \textbf{86}, 419–478 (2014).

\bibitem{GuehneToth2009} O.~G\"uhne and G.~T\'oth, Phys. Rep. \textbf{474}, 1–75 (2009).
\bibitem{James2001} D.~F.~V.~James, P.~G.~Kwiat, W.~J.~Munro, and A.~G.~White, Phys. Rev. A \textbf{64}, 052312 (2001).
\bibitem{ParisRehacek2004} M.~G.~A.~Paris and J.~\v{R}eh\'a\v{c}ek (eds.), \textit{Quantum State Estimation} (Springer, 2004).
\bibitem{BlumeKohout2010} R.~Blume-Kohout, New J. Phys. \textbf{12}, 043034 (2010).

\bibitem{Kaltenbaek2012} R.~Kaltenbaek \textit{et al.}, Exp. Astron. \textbf{34}, 123 (2012).

\bibitem{Kaltenbaek2023} R.~Kaltenbaek \textit{et al.}, Quantum Sci. Technol. \textbf{8}, 014007 (2023).

\bibitem{RomeroIsart2011} O.~Romero-Isart \textit{et al.}, Phys. Rev. A \textbf{83}, 013803 (2011).

\bibitem{Millen2020} J.~Millen and B.~A.~Stickler, Rep. Prog. Phys. \textbf{83}, 026401 (2020).

\bibitem{Predehl2012} K.~Predehl \textit{et al.}, Science \textbf{336}, 441 (2012).

\bibitem{Peterson1993}
J.~Peterson,
``Observations and modeling of background seismic noise,''
U.S. Geological Survey Open-File Report 93-322 (1993).

\bibitem{Harms2015}
J.~Harms,
``Terrestrial gravity fluctuations,''
Living Rev. Relativ. \textbf{18}, 3 (2015).


\bibitem{Kubo1962}
R.~Kubo, 
``Generalized cumulant expansion method,'' 
J. Phys. Soc. Jpn. \textbf{17}, 1100 (1962).

\bibitem{dephasingReview2017}
G.~de~Lange, Z.~H.~Wang, D.~Ristè, V.~V.~Dobrovitski, and R.~Hanson,
``Universal dynamical decoupling of a single solid-state spin from a spin bath,''
Science \textbf{330}, 60–63 (2010).

\bibitem{Kaltenbaek2016}
R.~Kaltenbaek \textit{et al.}, 
\textit{MAQRO: Testing the Quantum Superposition Principle in Space}, 
arXiv:1607.08649 (2016).


\end{thebibliography}

\begin{tikzpicture}
    \draw (0,0) -- (4,0);
\end{tikzpicture}

\newpage

\begin{center}
\Large
\textbf{Supplementary Materials}

\vspace{0.5cm}
\end{center}

\setcounter{section}{0}
\renewcommand{\thesection}{S\arabic{section}}
\renewcommand{\theequation}{S\arabic{equation}}

%%%%%%%%%%%%%%%%%%%%%%%%%%%%%%%%%%%%%%%%%%%%%%%%%%%%%%%%%%%%
\section{Mapping clock cross-spectra to proper-time noise}
%%%%%%%%%%%%%%%%%%%%%%%%%%%%%%%%%%%%%%%%%%%%%%%%%%%%%%%%%%%%

This section provides the detailed physical justification for the mapping between co-located optical-clock cross-spectra, proper-time fluctuations, and the common-mode dephasing rate that enters the gravitational-entanglement dynamics. It addresses the question of why frequency correlations measured by atomic clocks can be used to bound the platform-invariant timing noise relevant to spatially separated massive interferometers.

\subsection*{S1.1 Proper time in the weak-field limit} \label{sec:s1}

In the weak-field and nonrelativistic limit of general relativity, the line element can be written as
\begin{equation}
d\tau(t,\mathbf{x})
\simeq \left[1 + \frac{\Phi(t,\mathbf{x})}{c^2} - \frac{v^2}{2c^2}\right] dt,
\label{eq:S1_tau}
\end{equation}
where $\Phi(t,\mathbf{x})$ is the Newtonian gravitational potential and $v$ is the velocity of the system. To leading order, proper-time fluctuations are therefore determined by perturbations of the scalar potential,
\begin{equation}
\varepsilon(t,\mathbf{x})
\equiv \frac{d\tau}{dt} - 1
\simeq \frac{\delta \Phi(t,\mathbf{x})}{c^2}.
\label{eq:S1_eps}
\end{equation}
Thus any nonrelativistic system --- atoms in an optical lattice, trapped ions, levitated nanospheres, or mesoscopic interferometer paths --- experiences the same class of scalar perturbations $\delta \Phi(t,\mathbf{x})$. This universality motivates using high-precision clocks to bound $\varepsilon(t,\mathbf{x})$.

\subsection*{S1.2 From clock frequency noise to proper-time noise}

Optical clocks measure the fractional frequency deviation
\begin{equation}
y(t) = \frac{\delta \nu(t)}{\nu_0}
= \frac{d}{dt}\varepsilon(t),
\label{eq:S1_y}
\end{equation}
where $\nu_0$ is the nominal optical transition frequency and $\varepsilon(t)$ is defined in Eq.~\eqref{eq:S1_eps}. For two independent clocks $A$ and $B$ located in the same laboratory, the one-sided cross-spectral density $S_{y_A y_B}(f)$ captures the part of the frequency noise that is common to both devices. 

If the relevant fluctuations have spatial correlation length $L_{\mathrm{corr}}$ much larger than the dimension $L_{\mathrm{lab}}$ of the experimental setup (typically $L_{\mathrm{lab}}\sim 0.1$--$1$~m), then $\delta \Phi(t,\mathbf{x})$ can be treated as spatially uniform across all clocks and interferometers in that lab. In this regime, one can write
\begin{equation}
y_A(t) \simeq \dot{\varepsilon}(t),\qquad
y_B(t) \simeq \dot{\varepsilon}(t),
\end{equation}
so that the cross-spectrum $S_{y_A y_B}(f)$ directly measures the one-sided spectrum of $\dot{\varepsilon}(t)$, i.e.
\begin{equation}
S_{y_A y_B}(f)
= (2\pi f)^2 S_{\varepsilon}^{(\mathrm{com})}(f),
\label{eq:S1_mapping}
\end{equation}
where $S_{\varepsilon}^{(\mathrm{com})}(f)$ is the common-mode proper-time noise spectral density. Equation~\eqref{eq:S1_mapping} is the relation used in the main text.

\subsection*{S1.3 Applicability to spatially separated massive interferometers}

In proposed gravity-mediated entanglement tests, the two masses are typically separated by distances
\begin{equation}
d \sim 10^{-4}\text{--}10^{-3}\,\mathrm{m},
\end{equation}
whereas the dominant sources of low-frequency potential fluctuations --- seismic motion, 
building vibrations, acoustic noise, slow laser-frequency drift, environmental
gravity gradients --- have correlation lengths of order 
\(L_{\mathrm{corr}}\sim 1\text{--}100\,\mathrm{m}\) at frequencies \(10^{-4}\text{--}10^{-1}\,\mathrm{Hz}\), as observed in background-noise and Newtonian-noise correlation studies \cite{Peterson1993,Harms2015}.

Under the experimentally valid condition
\begin{equation}
d \ll L_{\mathrm{corr}},
\label{eq:S1_corr_condition}
\end{equation}
the proper-time processes at the two mass locations are practically identical,
\begin{equation}
\varepsilon_A(t) \simeq \varepsilon_B(t) \simeq \varepsilon(t).
\end{equation}
As a result, the ``platform-invariant'' proper-time noise probed by co-located optical clocks is precisely the common-mode timing noise that appears in the dephasing generator $\mathcal{D}[Z_A+Z_B]$ for the two-mass interferometer.

The physical content of this assumption is simply that long-wavelength metric perturbations act as a classical background affecting all systems in the same laboratory identically. Short-range inhomogeneities (e.g.\ gradients on the millimeter scale) are not constrained by clock correlations, but those contribute dominantly to local dephasing and are absorbed into $\gamma_{\mathrm{loc}}$.

\subsection*{S1.4 Relation to the Newtonian interaction phase}

The total phase accumulated by a two-mass system in an interferometric geometry can be expressed schematically as
\begin{equation}
\phi_{AB}(t)
= \frac{1}{\hbar}\int_0^t
\Big[V_{AB}^{(\mathrm{grav})} 
+ \delta V_A(t') + \delta V_B(t')\Big] dt',
\label{eq:S1_phi_ab}
\end{equation}
where $V_{AB}^{(\mathrm{grav})}$ is the Newtonian gravitational interaction energy, and $\delta V_{A,B}$ represent all additional time-dependent potential fluctuations at the two locations. 

The coherent entangling phase is generated by the geometry-dependent part of $V_{AB}^{(\mathrm{grav})}$ and yields the coupling rate $\chi$ used in the main text. The fluctuating contributions $\delta V_{A,B}(t)$ decompose into a common-mode part and a differential part. Under the correlation condition~\eqref{eq:S1_corr_condition}, the common-mode term has the same temporal structure $\varepsilon(t)$ that clocks measure, and therefore enters the dephasing generator as a collective Lindblad operator $Z_A+Z_B$ with rate $\gamma_{\mathrm{com}}$. The differential part contributes to the local dephasing rate $\gamma_{\mathrm{loc}}$.

In summary, the use of co-located clock cross-spectra to bound $\gamma_{\mathrm{com}}$ is justified whenever the mass separation is much smaller than the correlation length of the relevant potential fluctuations. In that experimentally motivated regime, Eq.~\eqref{eq:S1_mapping} yields a direct and physically transparent constraint on the common-mode proper-time sector.

%%%%%%%%%%%%%%%%%%%%%%%%%%%%%%%%%%%%%%%%%%%%%%%%%%%%%%%%%%%%
\section{Filter-function relation and common-mode dephasing rate}
%%%%%%%%%%%%%%%%%%%%%%%%%%%%%%%%%%%%%%%%%%%%%%%%%%%%%%%%%%%%

For completeness, we summarize the filter-function derivation that links proper-time noise spectra to the common-mode dephasing rate. The starting point is the coherence functional for a qubit undergoing pure dephasing due to a stochastic phase process $\phi(t)$,
\begin{equation}
\mathcal{V}(t)
= \left\langle
e^{i\phi(t)}
\right\rangle
= \exp\!\left[
-\frac{1}{2}\langle \phi^2(t)\rangle
\right],
\label{eq:S2_visibility}
\end{equation}
where the last equality holds for Gaussian stationary noise. For a system with Hamiltonian $H=\lambda Z\,\varepsilon(t)$, the phase is
\begin{equation}
\phi(t) = \lambda \int_0^t \varepsilon(t')\,dt'.
\end{equation}
Using the Wiener–Khinchin theorem for the autocorrelation of $\varepsilon(t)$ and its one-sided spectral density $S_{\varepsilon}(f)$, one obtains the standard filter-function relation
\begin{equation}
\mathcal{V}(t)
= \exp\!\left[
-\lambda^2\int_0^\infty 
S_{\varepsilon}(f)\,
\frac{\sin^2(\pi f t)}{(\pi f)^2} df
\right].
\label{eq:S2_filter}
\end{equation}

In the white-noise limit $S_{\varepsilon}(f)\simeq S_{\varepsilon}(0)$ over the relevant frequency band, the sine term can be coarse-grained using
\begin{equation}
\int_0^\infty 
\frac{\sin^2(\pi f t)}{(\pi f)^2} df
= \frac{t}{2},
\end{equation}
so that Eq.~\eqref{eq:S2_filter} reduces to
\begin{equation}
\mathcal{V}(t)
= \exp\!\left[-\gamma_\phi t\right],\qquad
\gamma_\phi = \frac{\lambda^2}{2} S_{\varepsilon}(0).
\label{eq:S2_gamma_phi}
\end{equation}
In the two-mass geometry considered in the main text, the common-mode coupling to proper-time fluctuations is $\lambda_{\mathrm{com}}=\omega_g/2$, where
\begin{equation}
\omega_g = \frac{G m^2}{\hbar}\frac{\Delta x}{r^2}.
\end{equation}
Inserting $\lambda_{\mathrm{com}}$ into Eq.~\eqref{eq:S2_gamma_phi} yields
\begin{equation}
\gamma_{\mathrm{com}}
= \frac{\omega_g^2}{8}\, S_{\varepsilon}(0),
\label{eq:S2_gamma_com}
\end{equation}
the expression used in the main text. When a full colored spectrum $S_{\varepsilon}(f)$ is available, the integral~\eqref{eq:S2_filter} can be evaluated numerically without the white-noise approximation.

%%%%%%%%%%%%%%%%%%%%%%%%%%%%%%%%%%%%%%%%%%%%%%%%%%%%%%%%%%%%
\section{LOCC structure of the common-mode sector}
%%%%%%%%%%%%%%%%%%%%%%%%%%%%%%%%%%%%%%%%%%%%%%%%%%%%%%%%%%%%

Here we briefly justify the LOCC structure of the common-mode dephasing channel. Consider the Markovian master equation
\begin{equation}
\frac{d\rho}{dt}
= \gamma_{\mathrm{com}}\,
\mathcal{D}[Z_A+Z_B]\rho,
\end{equation}
with Lindblad dissipator
\begin{equation}
\mathcal{D}[L]\rho
= L\rho L^\dagger
- \frac{1}{2}\{L^\dagger L,\rho\}.
\end{equation}
Because $Z_A+Z_B$ commutes with itself at all times and has a discrete spectrum, the corresponding quantum channel over a time interval $t$ can be written as a convex mixture of unitary phase rotations,
\begin{equation}
\Phi_{\mathrm{com}}(\rho)
= \int d\phi\, p_t(\phi)\,
\bigl(e^{i\phi Z_A}\otimes e^{i\phi Z_B}\bigr)\,
\rho\,
\bigl(e^{-i\phi Z_A}\otimes e^{-i\phi Z_B}\bigr),
\label{eq:S3_channel}
\end{equation}
where $p_t(\phi)$ is a Gaussian distribution whose variance grows linearly with $t$ and is fixed by $\gamma_{\mathrm{com}}$. This representation shows that the channel amounts to a random but perfectly correlated local phase applied to $A$ and $B$.

Since the operations $e^{i\phi Z_A}$ and $e^{i\phi Z_B}$ are local unitaries and the classical label $\phi$ is sampled from a classical distribution, the map $\Phi_{\mathrm{com}}$ lies in the LOCC (local operations and classical communication) class. As such, it cannot generate entanglement from separable states. Adding local dephasing terms $\mathcal{D}[Z_A]$ and $\mathcal{D}[Z_B]$ preserves this LOCC character. This is the sense in which the platform-invariant proper-time sector defines a classical channel.

%%%%%%%%%%%%%%%%%%%%%%%%%%%%%%%%%%%%%%%%%%%%%%%%%%%%%%%%%%%%
\section{Derivation of the CHSH feasibility condition}
%%%%%%%%%%%%%%%%%%%%%%%%%%%%%%%%%%%%%%%%%%%%%%%%%%%%%%%%%%%%

The entanglement feasibility condition used in the main text is derived from a standard Bell–CHSH visibility threshold. For a balanced two-qubit state subjected to dephasing at total rate $\Gamma_{\phi}$, the visibility of interference fringes has the form
\begin{equation}
V(t) = C\, e^{-2\Gamma_{\phi} t},
\label{eq:S4_visibility}
\end{equation}
where $C$ is the contrast in the absence of dephasing. The CHSH parameter is given by
\begin{equation}
S(t) = 2\sqrt{2}\,V(t),
\end{equation}
and a Bell violation requires $S(t)>2$, i.e.
\begin{equation}
V(t) > \frac{1}{\sqrt{2}}.
\end{equation}
Substituting Eq.~\eqref{eq:S4_visibility} into this inequality yields
\begin{equation}
C\, e^{-2\Gamma_{\phi} t}
> \frac{1}{\sqrt{2}},
\end{equation}
or equivalently
\begin{equation}
2\Gamma_{\phi} t 
< \ln(C\sqrt{2}).
\end{equation}
Writing $\Gamma_{\phi}=\gamma_{\mathrm{loc}}+\gamma_{\mathrm{com}}$ and rearranging terms gives
\begin{equation}
\Gamma_{\phi} t 
< \frac{1}{2}\ln(\sqrt{2}/C).
\label{eq:S4_deph_bound}
\end{equation}

On the other hand, the gravitational entangling phase is characterized by the dimensionless product
\begin{equation}
\chi t
= \left[\frac{G m^2}{\hbar}\frac{\Delta x^2}{r^3}\right] t,
\end{equation}
as defined in the main text. A necessary condition for observing a Bell violation in the presence of dephasing is that the entangling phase exceed the dephasing budget, that is
\begin{equation}
\chi t \gtrsim \Gamma_{\phi} t + \frac{1}{2}\ln(\sqrt{2}/C),
\label{eq:S4_feasibility}
\end{equation}
which is the feasibility inequality used to generate the plots in Sec.~VII of the main text. Equation~\eqref{eq:S4_feasibility} makes explicit how the requirements on $\chi$ depend on the contrast $C$ and the total dephasing rate.

%%%%%%%%%%%%%%%%%%%%%%%%%%%%%%%%%%%%%%%%%%%%%%%%%%%%%%%%%%%%
\section{Details of the feasibility plots and parameter extraction}
%%%%%%%%%%%%%%%%%%%%%%%%%%%%%%%%%%%%%%%%%%%%%%%%%%%%%%%%%%%%

For completeness, we outline how the numerical values entering the 
feasibility plots are obtained from the published parameters summarized in 
Table~2 of the main text.

For each platform (QGEM, MAQRO, levitated optomechanics) we proceed as follows.

\subsection*{S5.1 Geometric coupling}

Given the reported mass $m$, superposition size $\Delta x$, and separation $r$, we compute the gravitational coupling rate
\begin{equation}
\chi = \frac{G m^2}{\hbar}\frac{\Delta x^2}{r^3}.
\end{equation}
Multiplying by the coherence time $t$ listed in the table yields the 
dimensionless product $\chi t$, corresponding to the quantity defined in 
Eq.~(11) of the main text.

\subsection*{S5.2 Dephasing budget}

The local decoherence rate $\gamma_{\mathrm{loc}}$ is taken from the corresponding experimental or proposal paper. For levitated platforms it is dominated by gas collisions, blackbody emission, and laser phase noise; for MAQRO it is controlled by residual gas and thermal radiation in a cryogenic environment; for QGEM it includes environmental noise and trap decoherence. In each case, where a range is reported, we use representative mid-range values to generate the plotted points, and we treat the spread as an uncertainty band. 

The common-mode timing rate $\gamma_{\mathrm{com}}$ is obtained from Eq.~(4) of the main text 
 using $S_{\varepsilon}(0)\lesssim 10^{-28}\,\mathrm{Hz^{-1}}$, consistent with optical-clock cross-spectral measurements \cite{Oelker2019,Bothwell2022,Takamoto2020}. For all platforms considered, this yields
\begin{equation}
\gamma_{\mathrm{com}}\sim 10^{-12}\,\mathrm{s^{-1}},
\end{equation}
which is negligible compared to $\gamma_{\mathrm{loc}}$ and does not affect the ordering of points in the feasibility plots.

The total dephasing rate is then
\begin{equation}
\Gamma_{\phi} = \gamma_{\mathrm{loc}}+\gamma_{\mathrm{com}}
\simeq \gamma_{\mathrm{loc}},
\end{equation}
and the product $\Gamma_{\phi} t$ is plotted on the vertical axis of $Figure 1$ in the main text.

\subsection*{S5.3 Contrast and threshold line}

The contrast $C$ is taken from the reported or target interferometric visibility for each platform. Typical values are $C\simeq 0.6$--$0.8$, and we choose representative mid-range values consistent with the cited works. The threshold line in each panel of Figure 1 in the main text is then computed from Eq.~\eqref{eq:S4_feasibility},
\begin{equation}
\chi t = \Gamma_{\phi} t + \frac{1}{2}\ln(\sqrt{2}/C),
\end{equation}
so that points lying above the line correspond to parameter regimes where a Bell inequality violation is feasible in principle.

\subsection*{S5.4 Interpretation}

With these definitions, the feasibility plots summarize the relative positions of different architectures in a model-independent way. Levitated systems are pushed deep into the classical regime by large $\gamma_{\mathrm{loc}}$, MAQRO occupies a region where $\Gamma_{\phi} t$ is small but $\chi t$ is also small, and QGEM lies closest to the threshold because it combines relatively large $\chi$ with coherence times of order one second. The negligible contribution of $\gamma_{\mathrm{com}}$ confirms that the platform-invariant proper-time sector does not limit any of these experiments at current metrological sensitivities.

\vspace{0.4cm}

\noindent
\textit{Note:} All citation keys used here are assumed to refer to the same bibliography as in the main text.

\end{document}